\documentstyle[12pt]{article}
\vspace{6cm}

\title{\Large{
   Hadronic contributions to the anomalous magnetic moment of the muon
   by QCD model with infinite number of vector mesons}}

\author{ B.V.Geshkenbein, V.L.Morgunov \\
  \small{Institute for Theoretical and Experimental Physics} \\
  \small{RU-117279  Moscow  Russia} \\  }

\date{\small{September 14, 1994}}

\begin{document}
\maketitle
\abstract{
We computed the hadronic vacuum-polarization contributions to the muon
anomalous magnetic moment $a_{\mu}(hadr.)$ by using the QCD model
with infinite number of vector mesons [1,2]

The result is $a_{\mu}(hadr.) = 663(23) \times 10^{-10}$. }

\vspace{3cm}

\small{E-mail geshken@vxitep.itep.ru, morgunov@vxdesy.desy.de}

\vspace{3cm}
%\newpage

The largest uncertainties in the existing theoretical calculations of
the gyromagnetic ratio of the muon $a_{\mu} = (g-2)/2$ come from the
order - $\alpha^2$ the hadronic contributions to the photon vacuum
polarization. In the present paper we suggest the simple
method of the calculation of the hadronic vacuum polarization
contributions $a_{\mu}(hadr.)$ by using the QCD model with an infinite
number of vector mesons [1,2].

The value of $a_{\mu}(hadr.)$ can be written as [3]
\begin{equation}%Eq.1
   a_{\mu}(hadr.) = \frac{\alpha^2}{3\pi^2} \int\limits_{4m_{\pi}^{2}}^{\infty}
    ds K(s)R(s)/s
\end{equation}
where:
\begin{eqnarray}%Eq.2
   K(s) = x^{2}(1-x^{2}/2) + (1+x)^{2}(1+x^{-2}) \left[ \ln(1+x) - x +
   x^{2}/2 \right] + \nonumber \\
   + \frac{1+x}{1-x} x^{2} \ln x  \;\;\; ; \;\;\
   x=\frac{1-(1-4m_{\mu}^{2}/s)^{1/2}}{1+(1-4m_{\mu}^{2}/s)^{1/2}}
\end{eqnarray}
where: $m_{\mu}$ is the muon mass.

The function $R(s)$ is
\begin{equation}%Eq.3
  R(s) = \frac{\sigma^{H}(s)}{\sigma(e^{+} e^{-} \rightarrow \mu^{+} \mu^{-})}
\end{equation}
where $\sigma^{H}$ represents $\sigma(e^{+} e^{-} \rightarrow hadrons)$
and
\begin{equation}%Eq.4
\sigma(e^{+} e^{-} \rightarrow \mu^{+} \mu^{-}) = \frac{4\pi \alpha^{2}}{3 s}
\end{equation}
 Let us write the function $R(s)$ in the form:
\[
   R = R_{\rho} + R_{\omega} + R_{s} + R_c + R_b
\]
Where $R_{\rho}$ and $R_{\omega}$ are the contributions of $u$ and $d$
quarks in the state with isotopic spin $I=1$ ($\rho$ - family) and $I=0$
($\omega$ - family) and $R_s$, $R_c$, $R_b$ are the contributions of $s$
$(\varphi$ - family), $c$ ($J/\psi$ - family) and $b$ ($\Upsilon$ - family)
quarks respectively.

Let us consider for example $J/\psi$ family. In the approximation of
an infinite number of narrow resonances, having masses $M_{k}$
and electronic widths $\Gamma_{k}^{ee}$ , the function $R_{c}(s)$ has form:
\begin{equation}%Eq.5
    R_{c}(s) = \frac{9 \pi}{\alpha^2} \sum_{k=0}^{\infty} \Gamma_{k}^{ee}
    M_{k} \; \delta (s - M_{k}^{2} )
\end{equation}
Where $\alpha^{-1} = 137.0359895$ [4].

The contribution of $R_c$ in $a_{\mu}(hadr.)$ has the form:
\begin{equation}%Eq.6
  a_{\mu}^{c} = \frac{3}{\pi} \sum_{k=0}^{\infty} f(s_{k}) \;\; , \;\;
  f(s_k) = \frac{\Gamma_{k}^{ee} K(s_{k})}{M_{k}}
   \;\; , \;\; s_k = M_{k}^{2}
\end{equation}
If for $k > 5$ the total widths $\Gamma_{k}$ and masses $M_{k}$
of the vector mesons obey the conditions
\begin{equation}%Eq.7
   M_{k}^{2} - M_{k-1}^{2} \ll M_{k} \Gamma_{k} \ll M_{k}^{2}
\end{equation}
then for $k > 5$ the function $R_{c} (s)$ will be described by a smooth
curve and describes an experimental data.
All the formulae of the model [5] can be used.

We transform the sum in Eq.(5) into an integral by means of the
Euler-Maclaurin formula [6] beginning from $k = k_0 = 4$.
\begin{eqnarray}%Eq.8
  \sum_{k=4}^{\infty} f(s_{k}) = I + \frac{1}{2}f(s_{4}) -
      \frac{1}{12}f^{(1)}(s_{4})+ \frac{1}{720} f^{(3)}(s_{4}) - \nonumber \\
      - \frac{1}{30240}f^{(5)}(s_{4}) + \cdots
\end{eqnarray}
In (8) we have introduced the notations
\begin{equation}%Eq.9
f^{(l)}(s_{4}) = \frac{\partial^l f(s_k)}{\partial k^l}\mid_{k=4}
, \; I = \int \limits_{s_4}^{\infty} f(s_k) \frac{dk}{ds_k} ds_k
\end{equation}
In Ref. [1,2] we established a correspondence between the electronic
width of k-th resonance $\Gamma_{k}^{ee}$ and a derivative of the mass of
k-th resonance $M_{k}$ with respect to the number of this resonance
\begin{equation}%Eq.10
   \Gamma_{k}^{ee} =  \frac{2 \alpha^2}{9 \pi}  R_{c}^{PT}(s_k)
    \frac{dM_{k}}{dk}
\end{equation}
The function $R_{c}^{PT}$ includes all corrections on $\alpha_{s}$ in
perturbation theory (PT).

The term $1/12 f^{(1)}(s_4)$ in Eq.(8) is approximately equal to
\begin{equation}%Eq.11
    \frac{1}{12} f^{(1)}(s_4) = \frac{1}{24} \left[ f(s_5) - f(s_3) \right]
\end{equation}
This term is small (see table 2, additional term). The remaining terms
in (8) \\
$1/720 \; f^{(3)}(s_4) - 1/30240 \; f^{(5)}(s_4) + \cdots$ may be
omitted due to their smallness. The quantity of $a_{\mu}^{c}$ practically
does not change if $k_0 = 1,2,3$. We can not estimate of the term
$1/12 \; f^{(1)}(s_{k_{0}})$ if $k_0 = 0$ or $5$ .

Integral $I$ in Eq.(9), after the substitution (10), is equal to
\begin{equation}%Eq.12
I = \frac{\alpha^2}{9\pi} \int\limits_{s_4}^{\infty}
    \frac{K(s) R_{c}^{PT}(s) ds}{s}
\end{equation}
where $R_{c}^{PT}$ we use the formula [1,7]:
\begin{equation}%Eq.13
  R_{c}^{PT}(s) = R_{c}^{(0)} (s) {\cal D} (s)
\end{equation}
where
\begin{equation}%Eq.14
  R_{c}^{(0)} (s) = \frac{3}{2} Q_{c}^{2} v (3 - v^2)
    \; , \; v = \sqrt{1-\frac{4m_{c}^{2}}{s}} \; , \; Q_{c} = \frac{2}{3}
\end{equation}
$m_c = 1.30 \pm 0.05 \; GeV$, [2] and
\begin{equation}%Eq.15
   {\cal D} (v) =
   \frac{4\pi\alpha_s / 3v}{1 - \exp(-4\pi\alpha_s / 3v)} - \frac{1}{3}
   (\frac{\pi}{2} - \frac{3}{4\pi})(3+v) \alpha_s(s)
\end{equation}
In Eq.(13) we took into account the terms of the first order in $\alpha_s$
and "Coulomb" terms of all orders in $\alpha_s /v$ . At $v \rightarrow 1$
\begin{equation}%Eq.16
    {\cal D} (s) \rightarrow 1 + \frac{\alpha_s (s)}{\pi}
\end{equation}
Note that the function ${\cal D} (s)$ differs from (16) strongly in the region
of resonances (for example ${\cal D} (s_4) = 1.34$ for
$J/\psi$ family,  ${\cal D} (s_4)= 1.59$ for $\Upsilon$ family), hence it is
necessary to
take into account the "Coulomb" term.
For $\alpha_s(s)$ we used formulae [8] described the
evolution of $\alpha_s(s)$ taking into account effects of flavor
thresholds and the new result -
$\alpha_s (m_{Z}^{2}) = 0.125 \pm 0.005$ [9]. Within the framework of
modified minimal substruction $MS$ scheme [10,11] we obtained
$\Lambda^{(5)}_{\overline{MS}} = 0.31(8)$ ,
$\Lambda^{(4)}_{\overline{MS}} = 0.42(9)$,
$\Lambda^{(3)}_{\overline{MS}} = 0.46(9)$ .

Final formulae for $a_{\mu}^{c}$ has the form:
\begin{equation}%Eq.17
a_{\mu}^{c} = a_{\mu}^{c}(Reson.) + a_{\mu}^{c} (Int.) + a_{\mu}^{c} (Add.)
\end{equation}
Where:
\[
 a_{\mu}^{c}(Reson.) =  \frac{3}{\pi} \left[
     \sum_{k=0}^{3} \frac{\Gamma_{k}^{ee} K(s_k)}{M_k} +
       \frac{\Gamma_{4}^{ee} K(s_4)}{2 M_4} \right]
\]
\[
a_{\mu}^{c}(Int.) = \frac{\alpha^2}{3\pi^2} \int\limits_{s_4}^{\infty}
    \frac{K(s) R_{c}^{PT}(s) ds}{s}
\]
\[
a_{\mu}^{c}(Add.) =  \frac{1}{8 \pi} \left(
    \frac{\Gamma_{3}^{ee}}{ M_{3}} - \frac{\Gamma_{5}^{ee}}{ M_{5}}  \right)
\]
The formulae for the contribution of the $\Upsilon$ - family in
$a_{\mu}^{b}$ are derived if we replace the index $c$ by $b$
$(m_b = 4.54 \pm 0.02\; GeV)$ [2].
The results of the calculations are given in Tables 2,3.

The main contribution to $a_{\mu}(hadr.)$ comes from the vector
mesons consisting of light quarks. First we consider $\rho$ family
($Q^{2}_{\rho} = 1/2$).

Instead of Eq.(17) we have formulae:
\begin{equation}%Eq.18
    a_{\mu}^{\rho} = a_{\mu}^{\rho}(Reson.) + a_{\mu}^{\rho} (Int.) +
       a_{\mu}^{\rho} (Add.)
\end{equation}
Where:
\[
 a_{\mu}^{\rho}(Reson.) =  \frac{3}{\pi} \left[
     \frac{\Gamma_{0}^{ee} K(s_0)}{M_0} +
       \frac{\Gamma_{1}^{ee} K(s_1)}{2 M_1} \right]
\]
\[
a_{\mu}^{\rho}(Int.) = \frac{\alpha^2 Q^{2}_{\rho}}{3\pi^2}
    \int\limits_{s_1}^{\infty}
    \frac{K(s) R^{PT}(s) ds}{s} ; \;\;\; Q^{2}_{\rho} = 1/2
\]
\[
a_{\mu}^{\rho}(Add.) =  \frac{1}{8 \pi} \left(
    \frac{\Gamma_{0}^{ee}}{ M_{0}} - \frac{\Gamma_{2}^{ee}}{ M_{2}}  \right)
\]
In Eq.(18) the replacement of the summation by an integraction has been
carried out starting from $k=1$ , and all terms not written have been
ignored.

For $R^{PT}(s)$ we use here the formula [12,13] :
\begin{equation}%Eq.19
R^{PT}_{\rho}(s) = 3 Q^{2}_{\rho} \left[ 1 + \frac{\alpha_s(s)}{\pi} +
r_1 \left( \frac{\alpha_{s}(s)}{\pi} \right)^2 + r_2 \left(
\frac{\alpha_{s}(s)}{\pi}\right)^3 \right]
+ o(\alpha_{s}^{4}(s))
\end{equation}
Where:
\[ r_1 = 1.9857 - 0.1153 f
\]
\[
r_2 = -6.6368 -1.2001f -0.0052f^2 -1.2395
\frac{(\sum_f Q_f)^2}{3\sum_f Q_{f}^{2}}
\]

Similar equations are valid for $a_{\mu}^{\omega}$ ($Q^{2}_{\omega} =
1/18, \omega$ - family) and $a_{\mu}^{s}$ ($Q^{2}_{s} = 1/9,
\varphi$ - family).
Masses and electronic widths of the $\rho$, $\omega$ and $\varphi$ resonances
are presented in Table 1. Note that in the  $\omega$ case we have two
different sets for the describtion of the experimental results [14].
The difference for both cases in value $a_{\mu}^{\omega}$ is negligible.

The results of the calculations are given in Tables 2,3. Summing up the
contributions  of all families we get:
\begin{eqnarray}%Eq.20
   a_{\mu}(hadr) = 663(23) \times 10^{-10}
\end{eqnarray}
Note that the QCD model with an infinite number of vector mesons
makes possible the calculation of the hadronic contribution to the
anomalous magnetic momentum of the muon with the accuracy of $3 \%$.
Also this model made possible the calculation of the hadronic contribution
to the electromagnetic coupling constant $\alpha (q^{2})$ at $q^{2} =
m^{2}_{Z}$
with an accuracy better than $1 \%$ [15].

Ths result (20) is in good agreement with
$ a_{\mu}(hadr) = 707(6)(17) \times 10^{-10}$ [16] ,
$a_{\mu}(hadr) = 710(10.5)(4.9) \times 10^{-10}$ [17] and with earlier
works [18,19], obtained by integration formula (1) with the experimental
cross-section for $e^+ e^-$ annihilation into hadrons.

Summing up the contribution (20) the small rest hadronic contribution
$\Delta_{\mu}(hadr.) = -4.07(0.71) \times 10^{-10}$ [16] the QED
contribution $a_{\mu}(QED) = 1165846984(17)(28) \times 10^{-12}$ [20] and the
weak-interaction contribution $a_{\mu}(weak) = 19.5(0.1) \times 10^{-10}$
[21] we get:
\begin{equation}%Eq.23
   a_{\mu}(Theor.) = 11659148(23) \times 10^{-10}
\end{equation}
in good agreement with the experimental value [22]
\begin{equation}%Eq.24
   a_{\mu}(Exper.) = 11659230(80) \times 10^{-10}
\end{equation}

We thank L.B. Okun for useful discussions.

\section{References}
{\frenchspacing
\begin{tabbing}

1.~~~\=B.V.~Geshkenbein, Yad.Fiz. {\bf 51}, 1121 (1990).\\
\\
2.\>B.V.~Geshkenbein, V.L.~Morgunov, SSCL-Preprint-534, (1993) \\
\> (hep-ph/9407230). \\
\\
3.\>M.~Gourdin and E.de Rafael, Nucl. Phys. {\bf B 10} (1969) 667. \\
\>L.~Durand, Phys. Rev. {\bf 128}, (1962) 441. \\
\\
4.\>Particle Data Book, Phys. Rev. {\bf 45D} partII (1992) III.54. \\
\\
5.\>B.V.~Geshkenbein, Yad.Fiz. {\bf 49}, 1138 (1989).\\
\\
6.\>Handbook of Mathematical Functions, Edited by M.~Abramowitz and\\
\>I.A.~Stegun, 1964.\\
\\
7.\>V.A.~Novikov, et al. Phys. Rep. {\bf 41} (1979) 1. \\
\\
8.\>W.J.~Marciano, Phys. Rev. {\bf D29}, (1984) 580. \\
\\
9.\>V.A.~Novikov, L.B.~Okun, A.N.~Rozanov, M.I.~Vysotsky, Preprint \\
\>CERN-TH 7217/94 (1994), to be published in Phys. Lett.\\
\\
10.\>G.'t~Hooft, Nucl. Phys. {\bf B 61} (1973) 455. \\
\\
11.\>W.A.~Bardin, A.~Buras, D.~Duke and Muto, Phys. Rev. {\bf D18}, (1978)
3998. \\
\>A.~Buras, Rev.Mod.Phys. {\bf 52}, (1980) 199. \\
\\
12.\>S.G.~Gorishny,A.L.~Kataev and S.A.~Larin, Phys. Lett. {\bf B 252}, (1991)
114. \\
\\
13.\>L.R.~Surguladze, M.A.~Samuel, Phys. Rev. {\bf D66}, (1991) 560. \\
\\
14.\>A.B.~Clegg, A.~Donnachie, Preprint M-C-TH-93-21.\\
\\
15.\>B.V.~Geshkenbein, V.L.~Morgunov, ITEP-Preprint 49-94, (1994) \\
\> (hep-ph/9407228). \\
\\
16.\>T.~Kinoshita, B.~Nizic and Y.~Okamoto, Phys. Rev. {\bf D31}, (1985) 2108.
\\
\\
17.\>J.A.~Casas, G,.~Lopez and F.J.~Yinduran, Phys. Rev. {\bf D32}, (1985) 736.
\\
\\
18.\>J.~Calmet, S.~Narison, M.~Perrottet and E.~de Rafael, Phys. Lett. {\bf
61B}, (1976) 283. \\
\>Rev.Mod.Phys. {\bf 49}, (1977) 21. \\
\\
19.\>V.~Barger, W.F.~Long and M.G.~Olsson, Phys. Lett. {\bf 60B}, (1975) 89. \\
\\
20.\>T.~Kinoshita, Phys. Rev. {\bf D47}, (1993) 5013. \\
\\
21.\>R.~Jackiw, S.~Weinberg, Phys. Rev. {\bf D5}, (1972) 2473. \\
\>G.~Altarelli, N.~Cabibbo and L.~Maiani, Phys. Lett {\bf 40B}, (1972) 415. \\
\>I.~Bars, and M.~Yoshimura, Phys. Rev. {\bf D6}, (1972) 374. \\
\>K.~Fujikawa, B.W.~Lee and A.I.~Sanda, Phys. Rev. {\bf D6}, (1972) 2923. \\
\>W.A.~Bardeen, R.~Gastmans and B.E.~Lautrup, Nucl. Phys. {\bf B46}, (1972)
319. \\
\\
22.\>J.~Bailey et al., Phys. Lett. {\bf 68B}, (1977) 191. \\
\>F.J.M.~Farley, and E.~Picasso, Ann. Rev. Nucl. Sci. {\bf 29}, (1979) 243. \\
\\

\end{tabbing}
}

\begin{table}[t]

\begin{center}

Table 1. The values of masses and electronic widths of resonances.
\vspace*{0.2cm}

$u,d$ - $\rho$  family [4,14]
\vspace*{0.2cm}

\begin{tabular}{|l|l|l|l|}
\hline
&0&1&2 \\
\hline
$M_{i,Exp.}$ [GeV] & 0.7681(5) & 1.463(25) & 1.73(3) \\
\hline
$\Gamma_{i,Exp.}^{ee}$ [KeV] & 6.77(32) & 2.5(9) & 0.69(15) \\
\hline
\hline
\end{tabular}
\vspace*{0.2cm}

$u,d$ - $\omega$  family [4,14], two variants
\vspace*{0.2cm}

\begin{tabular}{|l|r|r|r|}
\hline
&0&1&2 \\
\hline
$M_{i,Exp.}$ [GeV] &0.78195(14)&1.44(7)&1.606(9) \\
\hline
$\Gamma_{i,Exp.}^{ee}$ [KeV] &0.60(2)& 0.150(38)&0.140(35) \\
\hline
\hline
$M_{i,Exp.}$ [GeV]  &0.78195(14)&1.628(14)& - \\
\hline
$\Gamma_{i,Exp.}^{ee}$ [KeV] &0.60(2)&0.37(10)& -  \\
\hline
\hline
\end{tabular}
\vspace*{0.2cm}

$s$ - $\varphi$  family [4]
\vspace*{0.2cm}

\begin{tabular}{|l|r|r|}
\hline
&0&1 \\
\hline
$M_{i,Exp.}$ [GeV] & 1.019413(8) & 1.70(2) \\
\hline
$\Gamma_{i,Exp.}^{ee}$ [KeV] & 1.37(5) & 0.70(18) \\
\hline
\hline
\end{tabular}
\vspace*{0.2cm}

$c$ - $J$/$\psi$  family [4]
\vspace*{0.2cm}

\begin{tabular}{|l|r|r|r|r|r|r|}
\hline
&0&1&2&3&4&5 \\
\hline
$M_{i,Exp.}$ [GeV]
& 3.09693(9) & 3.6860(1) & 3.7699(25) & 4.04(1) & 4.159(20) & 4.415(6) \\
\hline
$\Gamma_{i,Exp.}^{ee}$ [KeV]
& 5.36(29) & 2.14(21) & 0.26(4) & 0.75(15) & 0.77(23) & 0.47(10)  \\
\hline
\hline
\end{tabular}
\vspace*{0.2cm}

$b$ - $\Upsilon$  family [4]
\vspace*{0.2cm}

\begin{tabular}{|l|r|r|r|r|r|r|}
\hline
&0&1&2&3&4&5 \\
\hline
$M_{i,Exp.}$ [GeV]
& 9.46032(22) & 10.02330(31) & 10.3553(5) & 10.5800(35)  & 10.865(8)  &
11.019(8) \\
\hline $\Gamma_{i,Exp.}^{ee}[KeV] $ & 1.34(4) & 0.56(9) &
0.44(4) & 0.24(5) & 0.31(7) & 0.13(3)  \\
\hline
\hline
\end{tabular}

\vspace*{0.5cm}

Table 2. The contributions to $a_{\mu}(hadr)$
\vspace*{0.2cm}

\begin{tabular}{|l|r|r|r|r|r|}
\hline
&  $u,d$ - $\rho$   & $u,d$ - $\omega$ &  $s$ - $\varphi$
&   $c$ - $J$/$\psi$  & $b$ - $\Upsilon$  \\
\hline
Resonances & 479(22) & 40(1) & 44(2) & 8(0) & 0(0) \\
Integral   &  51(01)  & 6(0)  & 8(0)  & 6(0) & 0(0) \\
Add. term  &  19(00)  & 2(0)  & 0(0)  & 0(0) & 0(0) \\
\hline
           & 549(23) & 48(1) & 52(2) & 14(0) & 0(0) \\
\hline
\hline
\end{tabular}

\vspace*{0.5cm}

Table 3. Summarized result
\vspace*{0.2cm}

\begin{tabular}{|l|r|}
\hline
\hline
& $a_{\mu}(hadr)$ \\
\hline
$u,d$ -  $\rho$       &  549(23)  \\
$u,d$ -  $\omega$     &   48(01)  \\
$s$ ~~~- $\varphi$    &   52(02)  \\
$c$ ~~~- $J$/$\psi$   &   14(00)  \\
$b$ ~~~- $\Upsilon$   &    0(00)  \\
\hline
  $a_{\mu}(hadr)$   &  663(23)   \\
\hline
\hline
\end{tabular}
\end{center}
\end{table}
\end{document}